\documentclass[12pt]{iopart}

\usepackage{amssymb}
\usepackage{amsthm}
\usepackage{amsfonts}
\usepackage{algorithmic}
\usepackage{enumerate}
\usepackage{latexsym}
\usepackage{graphicx}
\usepackage{bm}
\usepackage{subfigure}
\usepackage[dvips]{color}

\begin{document}

\title[Orbital Magnetization]{Numerical Study of Disorder on the Orbital Magnetization in Two Dimensions}

\author{Si-Si Wang$^{1,2}$, Yan-Yang Zhang$^3$, Ji-Huan Guan$^{1,4}$, Yan Yu$^{1,4}$, Yang Xia$^{5,6}$ and Shu-Shen Li$^{1,2,7}$}
\address{$^1$ SKLSM, Institute of Semiconductors, Chinese Academy of Sciences, P.O. Box 912, Beijing 100083, China}
\address{$^2$ College of Materials Science and Opto-Electronic Technology, University of Chinese Academy of Sciences, Beijing 100049, China}
\address{$^3$ School of Physics and Electronic Engineering, Guangzhou University, 510006 Guangzhou, China}
\ead{yanyang@gzhu.edu.cn}
\address{$^4$ School of Physical Sciences, University of Chinese Academy of Sciences, Beijing 100049, China}
\address{$^5$ Microelectronic Instrument and Equipment Research Center, Institute of Microelectronics of Chinese Academy of Sciences, Beijing 100029, China}
\address{$^6$ School of Microelectronics, University of Chinese Academy of Sciences, Beijing 100049,
China}
\address{$^7$ Synergetic Innovation Center of Quantum Information and Quantum Physics, University of Science and Technology of China, Hefei, Anhui 230026, China}

\vspace{10pt}
\begin{indented}
\item{\today}
\end{indented}

\begin{abstract}
The modern theory of orbital magnetization (OM) was developed by using Wannier function method, which has a formalism similar with the Berry phase.
In this manuscript, we perform a numerical study on the fate of the OM under disorder, by using this method on the Haldane model in two dimensions, which can be tuned between a normal insulator or a Chern insulator at half filling.
The effects of increasing disorder on OM for both cases are simulated.
Energy renormalization shifts are observed in the weak disorder regime and topologically trivial case, which was predicted by a self-consistent $T$-matrix approximation. Besides this, two other phenomena can be seen. One is the localization trend of the band orbital magnetization. The other is the remarkable contribution from topological chiral states arising from nonzero Chern number or large value of integrated Berry curvature.
If the fermi energy is fixed at the gap center of the clean system, there is an enhancement of $|M|$ at the intermediate disorder, for both cases of normal and Chern insulators, which can be attributed to the disorder induced topological metal state before localization.
\end{abstract}

%
%
%
%
%

\section{Introduction}

The quantum theory of orbital magnetization (OM) was brought to the forefront by the end of the 20th century. This theory was first derived by using linear-response methods which was only used to calculate the OM changes instead of the OM itself\cite{Mauri1996,Mauri1996B,Mauri2002,Sebastiani2002,Sebastiani2001}. After 2005, a systematic quantum mechanical method was proposed to calculate the OM itself of crystalline insulators in the Wannier representation\cite{Marzari1997,Thonhauser2005,Resta2005}, which is consistent with semi-classical derivations\cite{Xiao2005,CFang2009}. It was further generalized to metals and Chern insulators\cite{Ceresoli2006,Marrazzo2016,Bianco}. These developments lead to modern theory of OM in solids\cite{Thonhauser2011,Aryasetiawan2019}.

This modern theory of OM is expressed as a momentum space Brillouin-zone integral function, with a formalism similar to that of Berry phase\cite{CFang2009,Thonhauser2011,Berry1984}.
Therefore, the OM manifests itself in an peculiar way in topological materials. A typical example is the large and energy dependent OM in the bulk gap as a result of chiral edge states associated with nonzero Chern number\cite{Ceresoli2006,Bianco}.

Experimentally, the OM can be measured by the Compton scattering of photons\cite{ComptonScattering,DeterminationOM0}, x-ray absorption or x-ray magnetic
circular dichroism spectroscopy\cite{DeterminationOM,DeterminationOM2}. For example, in the coexistence of spin and orbit magnetic moments, the orbital magnetic moment can be obtained from the total one by deducting the spin counterpart measured from the magnetic Compton scattering in terms of an applied magnetic field\cite{DeterminationOM0}. Several materials possessing remarkable bulk orbital magnetic moment have been recently proposed\cite{Nikolaev2014,CorrelatedElectron,RoleOfBerryPhase,Liu2019,WeylSemimetal} or observed\cite{DeterminationOM0,DeterminationOM,DeterminationOM2,DeterminationOM3}. It was also found that the OM plays an important role in the process of magnetization switching operation\cite{MagnetizationSwitching}.

Most researches of OM so far have been focused on clean crystals. It is well known that strong disorder will eventually induce localization\cite{Anderson1958,Hatsugai1999,YY2012,YY2013,Localization19B}.
Disorder also leads to rich phenomena in topological materials even in two dimensions, for example, the topological Anderson insulator\cite{YY2012,TAI,TAI2018} and the topological metal\cite{TopologicalMetal2,TopologicalMetal}.
The essential physics underlying these quantum transports is the ``translational'' motion of electrons from one terminal to another through the sample.
The OM, on the other hand, is related to the angular momentum of the circular motion, which can be further separated into
the itinerant circular, the local circular and the topological (boundary circular) contributions\cite{Thonhauser2005,Ceresoli2006,Bianco}.
The effects of disorder on the OM and its process towards localization are still open questions.
Based on a self-consistent $T$-matrix approximation, it was concluded that the effect of weak disorder in two dimensions is simply an energy renormalization, i.e., a shift of orbital magnetization function along the energy axis\cite{GBZhu2012}.

In this manuscript, we perform numerical studies of the OM $M$ in disordered two-dimensional (2D) systems, based on the Haldane model\cite{Haldane1988,Haldane2004} with a tunable Chern number $C$.
Starting from a clean system with Chern number $C=0$ or $1$, the development of fermi energy dependent $M$ with increasing disorder is investigated.
Meanwhile, the development of the intrinsic anomalous Hall conductance (AHC) $\sigma_{xy}^{\mathrm{int}}$ is also presented as a reference indication for disorder induced topological transition.
Based on these numerical results, we demonstrate that, although the self-consistent $T$-matrix approximation can capture some features of the band OM in the weak disorder regime, it cannot predict the localization trend, and the contribution associated with chiral states which are important when $\sigma_{xy}^{\mathrm{int}}\gg 0$. If the fermi energy $\mu$
is fixed at the gap center of the clean system, we find there is always a peak of $|M|$ at the intermediate disorder, in both cases of $C=0,1$, before the final localization at strong disorder. This may correspond to a disorder induced metal state with non-quantized
$\sigma_{xy}^{\mathrm{int}}$\cite{TopologicalMetal2,TopologicalMetal}.

This manuscript is organized as follows. In sections II and III, the details of the model and calculation methods are introduced. Then, the results for topologically trivial ($C=0$) and nontrivial ($C=1$) phases are represented in Sections IV and V, respectively.

\section{The Model }
\begin{figure}[htbp]
\includegraphics*[width=0.7\textwidth]{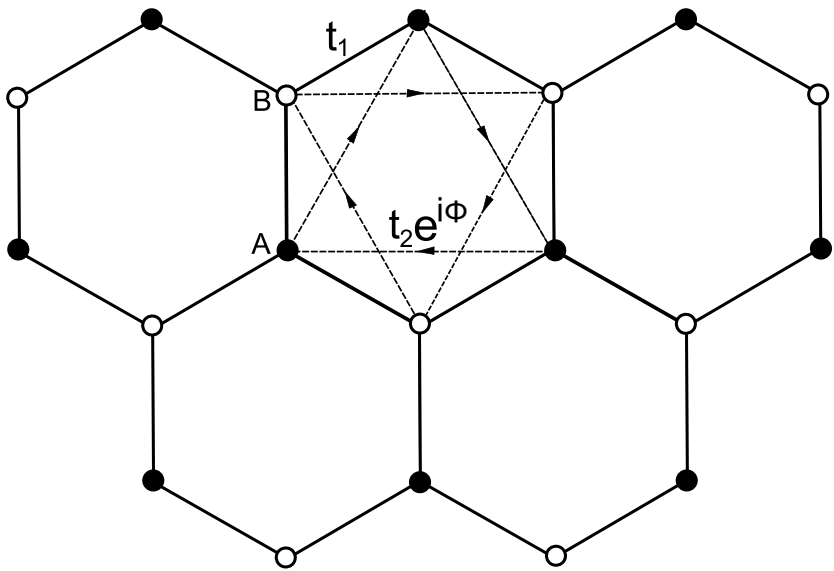}
\caption{ The honeycomb lattice of the Haldane model. Solid (open) points indicate the A (B) sublattice sites with onsite energy $+\Delta$ ($-\Delta$). $t_1$ represents the hopping amplitude of nearest-neighbor sites with different sublattice. $t_2e^{i\phi}$ represents the complex hopping amplitude of next nearest-neighbor sites with the same sublattice. Arrows on the dashed lines denote the positive sign of the phase $\phi$.  }
\label{structure}
\end{figure}

In order to incorporate the effects from topology, we adopt the Haldane model which can be tuned between a Chern insulator (quantum anomalous Hall effect) with $C=1$ and a normal insulator with $C=0$\cite{Haldane1988}. This is a tight binding model defined on a 2D honeycomb lattice,
with a real nearest-neighbor hopping $t_1$ (set to be 1 as the energy unit), complex next nearest-neighbor hoppings $t_2e^{\pm{i\phi}}$ and staggered potentials $\pm{\Delta}$,
where $\pm$ corresponds to A and B sublattices respectively, as illustrated in Fig. \ref{structure}.

Let $\bm{a_1},\bm{a_2},\bm{a_3}$ be the unit vectors from a site on sublattice A to its three nearest-neighbor sites on sublattice B, and $\bm{b_1}=\bm{a_2}-\bm{a_3},\bm{b_2}=\bm{a_3}-\bm{a_1},
\bm{b_3}=\bm{a_1}-\bm{a_2}$ be the next nearest-neighbor
vectors from this site to its three nearest-neighbor sites on sublattice A.
Now the Hamiltonian of Haldane model can be expressed in $k$-space as a $2\times 2$ matrix
\begin{equation}\label{Hk}
\eqalign{
 H(\bm{k})=2t_2\cos{\phi}\big(\sum_i\cos(\bm{k}\cdot\bm{b_i})\big)I+\\
 t_1\big(\sum_i\big[\cos(\bm{k}\cdot\bm{a_i})\sigma^1+
 \sin(\bm{k}\cdot\bm{a_i})\sigma^2\big]\big)+\\
 \big[\Delta-2t_2\sin{\phi}\big(\sum_i\sin(\bm{k}\cdot\bm{b_i})\big)\big]\sigma^3
 }
\end{equation}
where $\sigma^i$ are Pauli matrices acting on the space of sublattice. Nonzero $\phi$ breaks time-reversal symmetry and this can give rise to nonzero OM and Chern number\cite{Xiao2010,Thouless1982,Nakai2016}. This model's topological phase depends on the value of the parameter ratio $|\frac{\Delta}{t_2}|$. It is topologically nontrivial with Chern number $|C|=1$ when $|\frac{\Delta}{t_2}|<3\sqrt{3}\sin{\phi}$. The real space Hamiltonian can be obtained by an inverse Fourier transformation of Eq. (\ref{Hk}).

\begin{figure}[htbp]
\includegraphics*[width=0.9\textwidth]{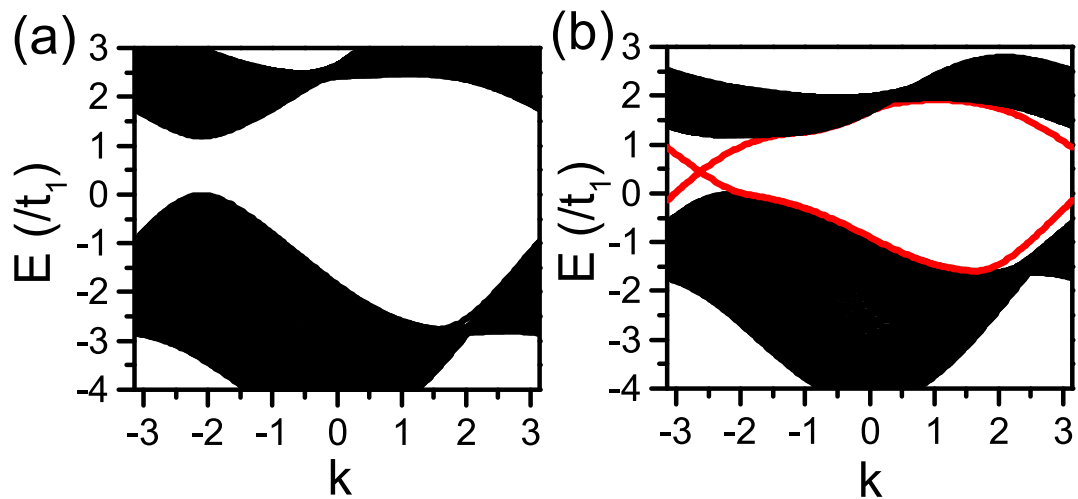}
\caption{ The band structure of the Haldane model in the quasi-one-dimensional ribbon geometry with width $N_y=100$ without aesny disorder. The two panels correspond to different onsite energy $\Delta$: (a) $\Delta=4.0$, (b) $\Delta=1.6$. The other model parameters are identical in both panels: $t_1=2.0, t_2=2/3, \phi=0.7\pi$. }
\label{disper}
\end{figure}

In Fig. \ref{disper}, we present the band structures of a ribbon with zigzag edges for two typical parameter settings, corresponding to $C=0$ in Panel (a) and $C=1$ in Panel (b), respectively. In both cases, the bulk gap region is around the energy range $(0,2)$. Red curves in Fig. \ref{disper}(b) are topological edge states traversing the bulk gap arising from $C=1$ of the valence band. Due to the topological origin, the existence of this pair of edge states is therefore robust for any kind of edge cut from the bulk crystal. Details of Haldane model¡¯s edge states associated with different edges are available in Ref. \cite{HaldaneEdges}. In the following, we will investigate the OM $M$ and its constituent parts [Eq. (\ref{magnet4})] for these two typical cases.

\section{The METHODS }

In a 2D crystalline solid, the quantum mechanical description of electronic OM, the orbital magnetic moment per unit volume (area in 2D discussed here), can be formulated in $k$-space as\cite{Thonhauser2005,Ceresoli2006,Thonhauser2011,Liu2019}
\begin{equation}\label{magnet}
\eqalign{
 M(\mu)&=M_{\mathrm{LC}}(\mu)+M_{\mathrm{IC}}(\mu)-2\mu N_{\mathrm{BC}}(\mu), \\
 M_{\mathrm{LC}}&=\frac{e}{2 c} \mathrm{Im}\sum_n\int_{\varepsilon_{nk}\leqslant\mu}
 \frac{\mathrm{d}\bm{k}}{(2\pi)^2}\,\langle\partial_{\bm{k}}u_{n\bm{k}}|
 \times H_{\bm{k}} |\partial_{\bm{k}}u_{n\bm{k}}\rangle \\
 M_{\mathrm{IC}}&=\frac{e}{2 c}\mathrm{Im}\sum_n\int_{\varepsilon_{nk}\leqslant\mu}
 \frac{\mathrm{d}\bm{k}}{(2\pi)^2}\,\varepsilon_{n\bm{k}}\langle\partial_{\bm{k}}u_{n\bm{k}}|
 \times|\partial_{\bm{k}}u_{n\bm{k}}\rangle \\
 N_{\mathrm{BC}}&=\frac{e}{2 c}\mathrm{Im}\sum_n
 \int_{\varepsilon_{nk}\leqslant\mu}\frac{\mathrm{d}\bm{k}}{(2\pi)^2}\, \langle\partial_{\bm{k}}u_{n\bm{k}}|
  \times|\partial_{\bm{k}}u_{n\bm{k}}\rangle\\
  &=\frac{e}{4\pi c}\mathcal{C},
  }
\end{equation}
where three terms correspond to the local circulation (LC), the itinerant circulation (IC) and the Berry curvature (BC) contributions, respectively.
Here $|u_{n\bm{k}}\rangle$ is the cell-periodic Bloch function, $c$ is the vacuum speed of light, and $\varepsilon_{nk}$ is the Bloch eigenvalue, so that $H_{\bm{k}}|u_{n\bm{k}}\rangle=\varepsilon_{n\bm{k}}|u_{n\bm{k}}\rangle$.
All summations in Eq. (\ref{magnet}) are over
occupied bands $n$ up to the fermi energy $\mu$. In the third term, $N_{\mathrm{BC}}$ is proportional to the (intrinsic) AHC $\sigma_{xy}^{\mathrm{int}}$ as \cite{RoleOfBerryPhase,Xiao2010}
\begin{equation}\label{EqHallConductivity}
\sigma_{xy}^{\mathrm{int}}=\frac{e^2}{h}\mathcal{C}=\frac{2ec}{\hbar} N_{\mathrm{BC}}.
\end{equation}
Here, the dimensionless number $\mathcal{C}$ is quantized as the Chern number $C$ (the topological invariant) when $\mu$ is in the gap, but may not be quantized if $\mu$ is not in the gap.
When $\mathcal{C}\neq 0$, this term corresponds to the contribution from the magnetic moment of chiral edge states\cite{Ceresoli2006}.

In numerical calculations, the derivatives $|\partial_{\alpha}u_{n\bm{k}}\rangle \equiv |\partial_{k_{\alpha}}u_{n\bm{k}}\rangle$ in Eq. (\ref{magnet}) have to be evaluated on a mesh of discretized Brillouin zone. However, this cannot be done by a simple finite-deference, since the gauges of wavefunctions
$| u_{n\bm{k}}\rangle$ on neighboring grid points cannot be fixed.
Instead, we use the discretized covariant derivative \cite{Marzari1997,Ceresoli2006,Souza2004}
\begin{equation}\label{EqCovariantDerivative}
|\tilde{\partial}_\alpha u_{n\bm{k}}\rangle,
\end{equation}
which involves linear combinations of occupied states under $\mu$, and a local gauge fixing around a definite grid point. See Appendix A of Ref. \cite{Ceresoli2006} for details. This definition guarantees itself to be numerically gauge invariant. Let us define\cite{Ceresoli2006,Souza2008}
\begin{equation}\label{fgh2}
\eqalign{
\tilde{f}_{\bm{k},i}&=\frac{1}{v}\epsilon_{ijl}q_{i}\sum_n\mathrm{Im}
               \langle\tilde{\partial}_{j} u_{nk}|\tilde{\partial}_{l}u_{nk}\rangle\\
\tilde{g}_{\bm{k},i}&=\frac{1}{v}\epsilon_{ijl}q_{i}\sum_n\mathrm{Im}
               \langle\tilde{\partial}_{j} u_{nk}|H_{\bm{k}}|
               \tilde{\partial}_{l}u_{nk}\rangle\\
\tilde{h}_{\bm{k},i}&=\frac{1}{v}\epsilon_{ijl}q_{i}\sum_{nm}
               \varepsilon_{nm\bm{k}}\mathrm{Im}
               \langle\tilde{\partial}_{j} u_{mk}|
               \tilde{\partial}_{l} u_{nk}\rangle\\
}
\end{equation}
where $q_{i}$ represents the primitive reciprocal vectors of the discretized $\bm{k}$ mesh in the $i$-th direction, and $v$ denotes the volume of the unit cell in the mesh. Now Eq. (\ref{magnet}) can be transformed into another form as \cite{Ceresoli2006,Souza2008}:
\begin{eqnarray}\label{magnet2}
\eqalign{
 &M_{\mathrm{LC}}=\frac{-1}{2c(2\pi)^2}\int_{\mathrm{BZ}}\mathrm{d}\bm{k}
   \tilde{g}_{\bm{k}},  \\
 &M_{\mathrm{IC}}=\frac{-1}{2c(2\pi)^2}\int_{\mathrm{BZ}}\mathrm{d}\bm{k}
   \tilde{h}_{\bm{k}},   \\
 &N_{\mathrm{BC}}=\frac{1}{2c(2\pi)^2}\int_{\mathrm{BZ}}\mathrm{d}\bm{k}
   \tilde{f}_{\bm{k}},
}
\end{eqnarray}
where $\tilde{f}_{\bm{k}}\equiv \tilde{f}_{\bm{k},z}$ (similarly for $\tilde{g}$ and $\tilde{h}$) is the only nonzero component for a 2D crystal.
Besides the numerical computability of $|\tilde{\partial}_\alpha u_{n\bm{k}}\rangle$, another merit of Eqs. (\ref{fgh2}) and (\ref{magnet2}) is that now the components $M_{\mathrm{LC}}$ and $M_{\mathrm{IC}}$ are separately gauge invariant, even in the multi-band case\cite{Ceresoli2006}.

Gauge invariant quantities like those defined in (\ref{magnet2}) should be potentially observable\cite{Xiao2010,Scott1962,Hugu1971}. As stated in Eq. (\ref{EqHallConductivity}), $N_{\mathrm{BC}}$ is proportional to the AHC $\sigma_{xy}^{\mathrm{int}}$\cite{Xiao2010,QAHRMP}. As for the other two terms, we adopt an alternate combination defined in Ref. \cite{Souza2008} as
\begin{equation} \label{magnet4}
 \eqalign{
 M(\mu)&=M_{\mathrm{SR}}^{(\mathrm{I})}(\mu)+\Delta M_0 (\mu)-2{\mu}N_{\mathrm{BC}}(\mu),\\
 M_{\mathrm{SR}}^{(\mathrm{I})}&=M_{\mathrm{LC}}-M_{\mathrm{IC}},\\
 \Delta{M_0}&=2 M_{\mathrm{IC}}.
 }
\end{equation}
Here, $M_{\mathrm{SR}}^{(\mathrm{I})}$ is found to be proportional to the differential absorption of right and left circular polarized light verified by the $\textit{f}$-sum rule\cite{PM1998} which corresponds to the self-rotation of the WFs, namely, the intraorbital part of OM $M$. On the other hand, $\Delta M_0$ is the interorbital part which corresponds to a boundary current circulation\cite{Xiao2006},

The AHC $\sigma_{xy}^{\mathrm{int}}$ and OM $M$ are results of an integration over energy up to $\mu$. It is also insightful to investigate their densities with respect to energy\cite{Yang2011,Qiao2016}. Specifically, we define
\begin{eqnarray}\label{cdensity}
\rho_{\sigma}(\mu)=\lim\limits_{\triangle{E}\rightarrow 0}\frac{\sigma_{xy}^{\mathrm{int}}(\mu+\triangle{E})-\sigma_{xy}^{\mathrm{int}}(\mu)}{\triangle{E}}
\end{eqnarray}
and
\begin{eqnarray}\label{mdensity}
\rho_{M}(\mu)=\lim\limits_{\triangle{E}\rightarrow 0}\frac{M(\mu+\triangle{E})-M(\mu)}{\triangle{E}}.
\end{eqnarray}

In the following, we will focus on how $M$ and its constituent components defined in Eq. (\ref{magnet4}) change with respect to disorder strength. The disorder is simulated as a random potential $U(i)$ on each site $i$, where $U(i)$ is uniformly distributed within the interval $(-\frac{W}{2},\frac{W}{2})$ , with $W$ the disorder strength. To be compatible with the $k$ space formalism introduced above\cite{Thonhauser2005,Bianco,Thonhauser2011}, we use a disordered sample with size $L_x \times L_y$ as a supercell (repeating unit) of a superlattice, so that the translation symmetry can be restored on a larger scale, and the results are free from artificial effects from existing edges. This is equivalent to twisted boundary conditions on both directions\cite{Hatsugai1999,YY2012,Xiao2010,Sheng1997,Essin2007,Song2016}. Due to the presence of disorder fluctuation, each data point in the following will be an average over disorder ensembles, with typically 1000 random configurations.

\section{Results and Discussion}

\subsection{Topologically Trivial Phase}

\begin{figure}[htbp]
\includegraphics*[width=0.9\textwidth]{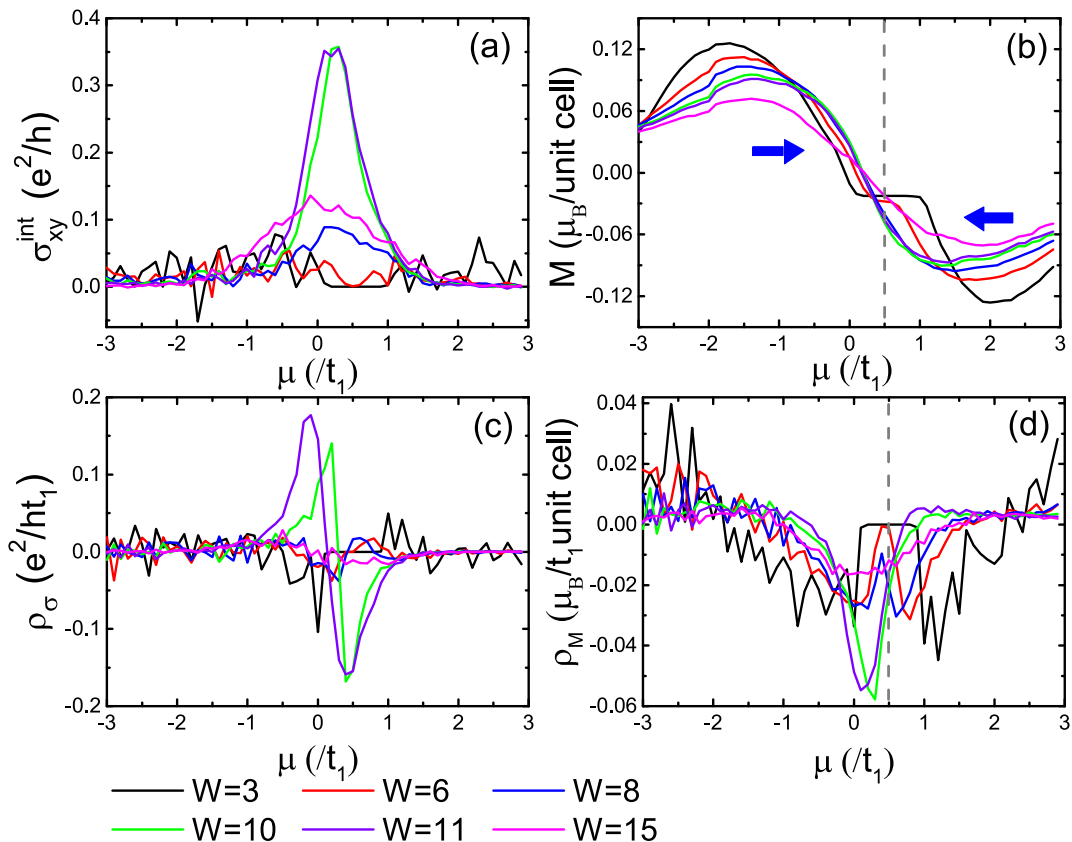}
\caption{ The fermi energy dependences for the topologically trivial case corresponding to Fig. \ref{disper} (a). The AHC $\sigma_{xy}^{\mathrm{int}}$  [Panel (a)] and OM $M$ [Panel (b)] as functions of fermi energy $\mu$, for different disorder strength $W$. Panels (c) and (d) are their energy densities, respectively. Each data point is an average over 1000 disorder configurations with the supercell size $L_x=L_y=L=6$. The blue arrows indicate the shift direction of the $M(\mu)$ profiles, as predicted in \cite{GBZhu2012}. The dashed lines in (b) and (d) indicate the location of gap center in the clean limit.}
\label{phi=0.7C=0}
\end{figure}

Let us start from a topologically trivial phase as shown in Fig. \ref{disper} (a), which is a normal insulator at half filling with Chern number $C=0$. In Fig. \ref{phi=0.7C=0} (a), we plot the AHC $\sigma_{xy}^{\mathrm{int}}$ as a function of fermi energy $\mu$, for different disorder strengths $W$. At the weakest disorder ($W=3$, black line), the energy interval $\mu\in(0,2)$ with $\sigma_{xy}^{\mathrm{int}}=0$
corresponds to the energy gap. With increasing disorder, this gap shrinks to zero at around $W=6$. After that, $\sigma_{xy}^{\mathrm{int}}$ grows remarkably to a considerable but non-quantized value $\sim 0.35\frac{e^2}{h}$ at $W\sim 10$, before the localization ($\sigma_{xy}^{\mathrm{int}}\sim0$) at strong disorder $W=15$. Such an emergence of nonzero $\sigma_{xy}^{\mathrm{int}}$ under increasing disorder is not rare in systems without particle-hole symmetry, and can be attributed to disorder induced band inversion \cite{YY2012,YY2013,TopologicalMetal,StatTI}. The non-quantization of $\sigma_{xy}^{\mathrm{int}}$ suggests that the system is in the metallic state, similar to that before the appearance of topological Anderson insulator\cite{YY2013,TopologicalMetal2,TopologicalMetal}. However, since a stable energy gap or mobility gap cannot be formed before another disorder induced band inversion into a trivial Anderson insulator at strong disorder\cite{Hatsugai1999,YY2012,YY2013}, this system cannot develop into a topological Anderson insulator with a quantized topological invariant (Chern number here).

The development of OM $M(\mu)$ under disorder is the main focus of this work. The numerical results are presented in Fig. \ref{phi=0.7C=0} (b). The first observation is that $M$ has opposite signs for the valance and conduction bands respectively. At weak disorder $W=3$, they are separated by the gap with constant $M$.
In a previous work based on the self-consistent $T$-matrix approximation for weak disorder, it has been predicted that the effect of weak disorder on $M(\mu)$ is just an energy renormalization, i.e., a global shift of $M(\mu)$ profile along the $\mu$ axis\cite{GBZhu2012}. With the increasing of disorder strength $W$ shown in Fig. \ref{phi=0.7C=0} (b), as predicted, the $M(\mu)$ profiles associated with the valance band (with $M>0$) and the conduction band (with $M<0$) do shift along the energy axis, but in opposite directions respectively (illustrated by blue arrows). An important feature that has not been captured in the self-consistent $T$-matrix approximation is the reduction of the magnitude $|M|$ with increasing $W$ for most of the band ranges, which corresponds to the localization tendency of the orbital motion.

The opposite directional shifts of $M(E)$ for conduction and valence bands at weak disorder can be attributed to the energy renormalization from disorder. Consider a generic two-band model
\begin{equation}
H_2=h_x\sigma_x+h_y\sigma_y+h_z\sigma_z.\label{EqTwoBand}
\end{equation}
The effect of non-magnetic disorder to the this model can be calculated within the first Born approximation as a self energy $\Sigma$. Its real part is\cite{Beenakker}
\begin{equation}\label{EqReSigma}
\mathrm{Re}\Sigma \propto W^2 \mathrm{Re} \int \frac{\epsilon - h_x\sigma_x - h_y\sigma_y-h_z\sigma_z }{\epsilon^2-h_x^2 - h_y^2 - h_z^2 } d^3h,
\end{equation}
where $\epsilon$ is the eigen-energy, and it is $\epsilon^+>0$ ($\epsilon^-<0$) for the conduction (valance) band. Expressed in terms of $\sigma_i$, $\mathrm{Re}\Sigma$ is also a $2\times2$ matrix.
Its diagonal elements
\begin{equation}\label{EqReSigmaDiag}
\propto W^2 \int \frac{\epsilon^{\pm} }{(\epsilon^{\pm})^2-h_x^2 - h_y^2 - h_z^2 } d^3h,
\end{equation}
will contribute to the energy renormalization, i.e., band shifts, with opposite signs for conduction and valence bands respectively. This approaching of two topologically trivial bands at weak disorder can also be understood in a simpler context as a second perturbation\cite{YY2012}, which plays an important role in forming the topological Anderson insulator.

The energy densities associated with $\sigma_{xy}^{\mathrm{int}}$ and $M$ are presented in Fig. \ref{phi=0.7C=0} (c) and (d). Despite strong fluctuations, some important information can still be drawn. For the AHC density in Fig. \ref{phi=0.7C=0} (c), the sharp peaks ($M>0$) and valleys ($M<0$) at $W\sim 10$ correspond to the creation of topological charges (Chern numbers) with opposite signs soon after the disorder induced band inversion\cite{Hatsugai1999,YY2012,Xiao2010}. This picture confirms again the origin of the appearance of $\sigma_{xy}^{\mathrm{int}}$ peaks around $W\sim 10$ in Fig. \ref{phi=0.7C=0} (a). As another result, the chiral edge states associated with these nonzero topological charges give rise to remarkable contributions to $M$. This is reflected by the valleys with largest $|\rho_{M}|$ in Fig. \ref{phi=0.7C=0}(d), which also appear at $W\sim 10$.

\begin{figure}[htbp]
\includegraphics*[width=0.6\textwidth]{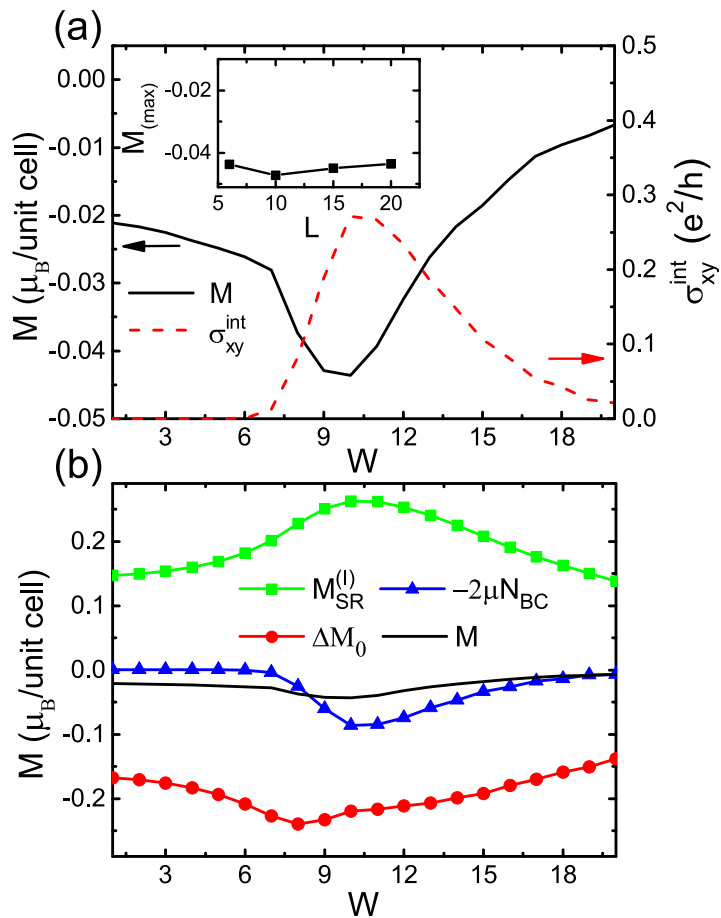}
\caption{ Developments of the normal insulator under disorder $W$, by fixing the fermi energy at $\mu=1$ [indicated as dashed line in Fig. \ref{phi=0.7C=0} (b) and (d)]. (a) The OM $M$ (black line) and anomalous Hall conductance $\sigma_{xy}^{\mathrm{int}}$ (red dashed line). The inset is the peak value $M_{(\mathrm{max})}$ versus the linear size $L$ of supercells.
(b) The OM $M$ (black line) and its constituent components (line with symbols) defined in Eq. (\ref{magnet4}).}

\label{fgh-C=0}
\end{figure}

After an overall view of $M(\mu)$, now we concentrate on the OM of an insulator\cite{Thonhauser2005,Bianco}, by fixing the fermi energy $\mu$ in the bulk gap center. In Fig. \ref{fgh-C=0} (a), we plot $M$ (black solid line) and $\sigma^{\mathrm{int}}_{xy}$ (red dashed line) as functions of disorder strength $W$, at the fermi energy $\mu=1$ [indicated as the dashed line in Fig. \ref{phi=0.7C=0} (c) and (d)], which is near the gap center in the clean limit.
The AHC $\sigma^{\mathrm{int}}_{xy}$ is identically zero until the band closing at $W\sim 6$. Notice that $M$ is changing during this process. This is not surprising since Chern number (here 0) is a topological invariant of the band while OM is not, and any distortion of the band (e.g., from disorder) may
influence the value of OM even when $\mu$ is in the gap.
With the appearance of nonzero $\sigma^{\mathrm{int}}_{xy}$, $|M|$ starts to increase more quickly. The magnitudes of both quantities arrive at the maximum value $|M|_{\mathrm{max}}$ and $\sigma^{\mathrm{int}}_{xy,\mathrm{max}}$ at the intermediate disorder around $W\sim 10$. From the size dependence of $|M|_{\mathrm{max}}$ in the inset of Fig. \ref{fgh-C=0} (a), it can be seen that this nonzero magnetization is expected to persist to the thermodynamic limit. This enhancement of orbital magnetic moment at intermediate disorder reflects the emergence of the metal state from another respect. Similarly, magnetic impurities was also found to induce remarkable OM in a Rashba electron gas\cite{MagneticImpurity}. After this peak at disorder $W\sim 10$, all electronic motions go towards a final localization $M,\sigma^{\mathrm{int}}_{xy}\sim 0$ in the strong disorder limit.

In order to obtain more intuitions, we scrutinize the behaviors of three constituent components of $M$: $M_{\mathrm{SR}}^{(\mathrm{I})}$, $\Delta{M_0}$ and $-2{\mu}N_{\mathrm{BC}}$ defined in Eq. (\ref{magnet4}). In Fig. \ref{fgh-C=0} (b), they are also presented as functions of disorder strength $W$, at the fermi energy $\mu=1$ near the gap center of the clean limit. Due to the vanishing of Chern number, the contribution from $-2{\mu}N_{\mathrm{BC}}$ term (blue line with triangular dots) is small, and is actually zero before the band touching at $W\sim 7$. However, the components $M_{\mathrm{SR}}^{(\mathrm{I})}$ (green line with square dots) and $\Delta{M_0}$ (red line with circle dots) are almost one order of magnitude larger than $M$ itself (black line), but with opposite signs. This cancelation makes the magnitude of total magnetization $M$ rather small. This feature is consistent with previous analytical results\cite{CFang2009,GBZhu2012}.

\subsection{Topologically Non-Trivial Phase}

\begin{figure}[htbp]
\includegraphics*[width=0.9\textwidth]{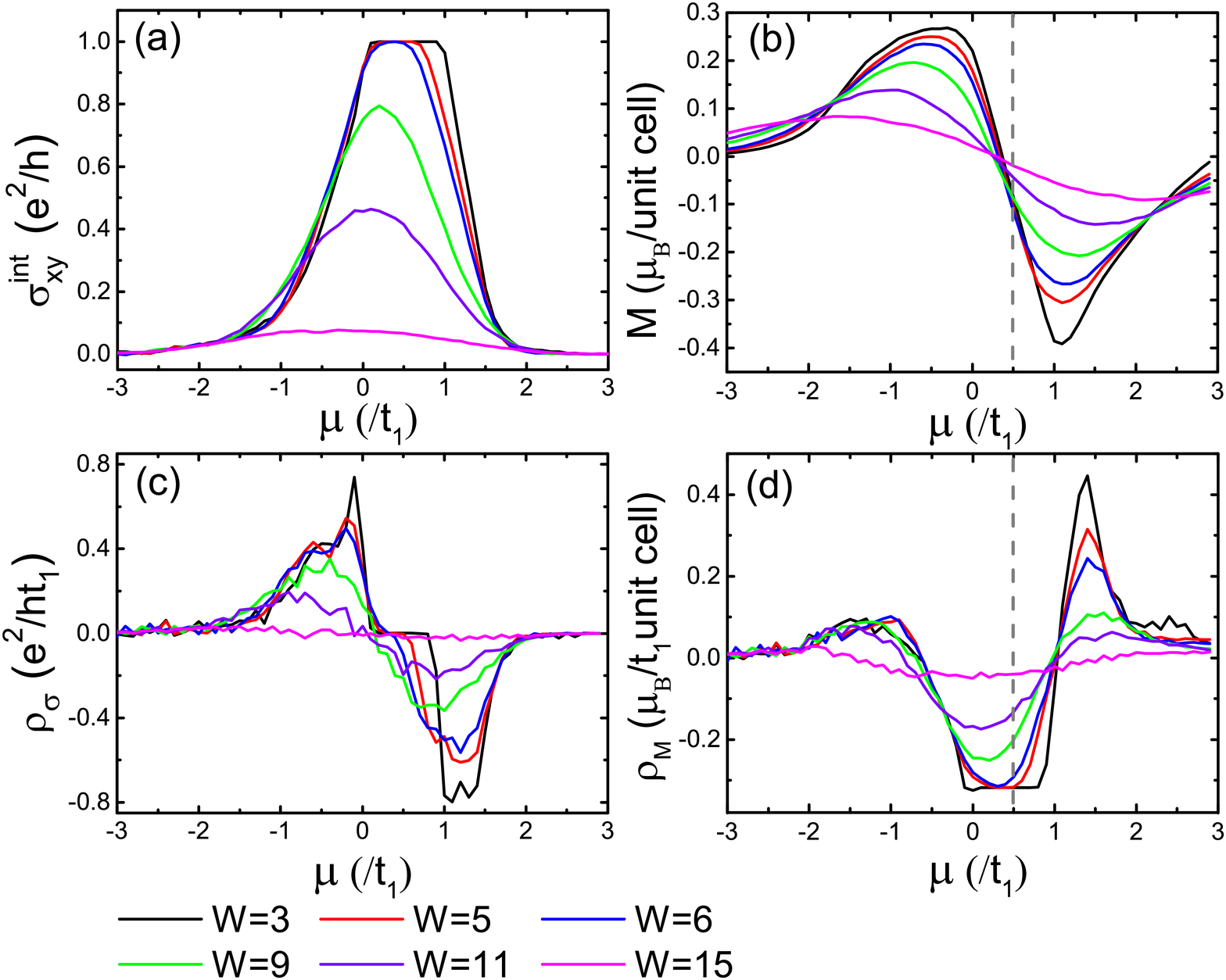}
\caption{ Similar to Fig. \ref{phi=0.7C=0}, but corresponding to the topologically nontrivial case shown in Fig. \ref{disper} (b).}
\label{Figphi=0.7C=1}
\end{figure}

Now we turn to the case of a Chern insulator at half filling, with the band structure as presented in
Fig. \ref{disper} (b).
The development of AHC $\sigma_{xy}^{\mathrm{int}}$ are plotted in Fig. \ref{Figphi=0.7C=1} (c). The $C=1$ plateau around $\mu\in(0,2)$ can be clearly seen in the weak disorder regime, reflecting the robust edge states in the bulk gap. With increasing disorder, the width of this plateau shrinks and finally results in a collapse after $W>6$.
The associated AHC densities $\rho_{\sigma}(\mu)$ for different disorder strengths are presented in Fig. \ref{Figphi=0.7C=1} (c). At the weakest disorder ($W=3$, black line), $\rho_{\sigma}(E)$ consists of one peak and one valley separated by a horizontal line with $\rho_{\sigma}(E)=0$, which correspond to the valance band with positive Chern number, the conduction band with negative Chern number, and the bulk gap, respectively. With increasing disorder, the gap shrinks and nonzero Chern numbers annihilate after the band touching. This is a well known process of Anderson localization for a Chern insulator at strong disorder, which also corresponds to a disorder induced band inversion\cite{Hatsugai1999,YY2012,Song2016,SS2019}.

The development of $M(\mu)$ under increasing disorder is presented in
Fig. \ref{Figphi=0.7C=1}(b). At the weakest disorder $W=3$ (black line), similar to the topologically trivial case, $M$ also possesses opposite signs in the valence and conduction bands respectively. Now in the gap region, $\mu\in(0,2)$, $M(\mu)$ is linearly decreasing instead of constant as in Fig. \ref{phi=0.7C=0} (b). This originates from the chiral edge states in the bulk gap, so that\cite{Ceresoli2006}
\begin{equation}\label{EqMEdgeStates}
\frac{dM}{d\mu}=-\frac{C}{2\pi c}.
\end{equation}
This is a direct consequence from the last term of Eq.  (\ref{magnet4}),
which is the only energy-dependent contribution when the fermi energy is in the bulk gap. From its energy density $\rho_M$ [black line in Fig. \ref{phi=0.7C=0} (d)], we can see that
dominating contributions are indeed from the bulk gap around $(0,2)$ and nearby band edges.
In other words, in the topologically nontrivial case, Berry curvature related chiral states play an important role in the orbital magnetization.

With increasing disorder, this linear region of $M$ shrinks gradually, due to the narrowing of the bulk gap. Meanwhile, the magnitudes of band orbital magnetization $M$ decrease almost monotonically in most of the \emph{band} region, as a result of the localization tendency.
Now, due to the strong modulation of chiral edge states pinned around the gap region, these $M(\mu)$ profiles associated with both bands do not exhibit prominent global shifts along the energy axis. This is different from the previous case with $C=0$ shown in Fig. \ref{phi=0.7C=0} (b), and also different from that predicted from the self-consistent $T$-matrix approximation\cite{GBZhu2012}. Therefore, the development of most band orbital magnetization $M$ under disorder in Fig. \ref{Figphi=0.7C=1}(b) looks simpler than that in Fig. \ref{phi=0.7C=0}: just a monotonic decreasing of magnitudes towards localization $|M|=0$ in strong disorder limit.

\begin{figure}[htbp]
\includegraphics*[width=0.6\textwidth]{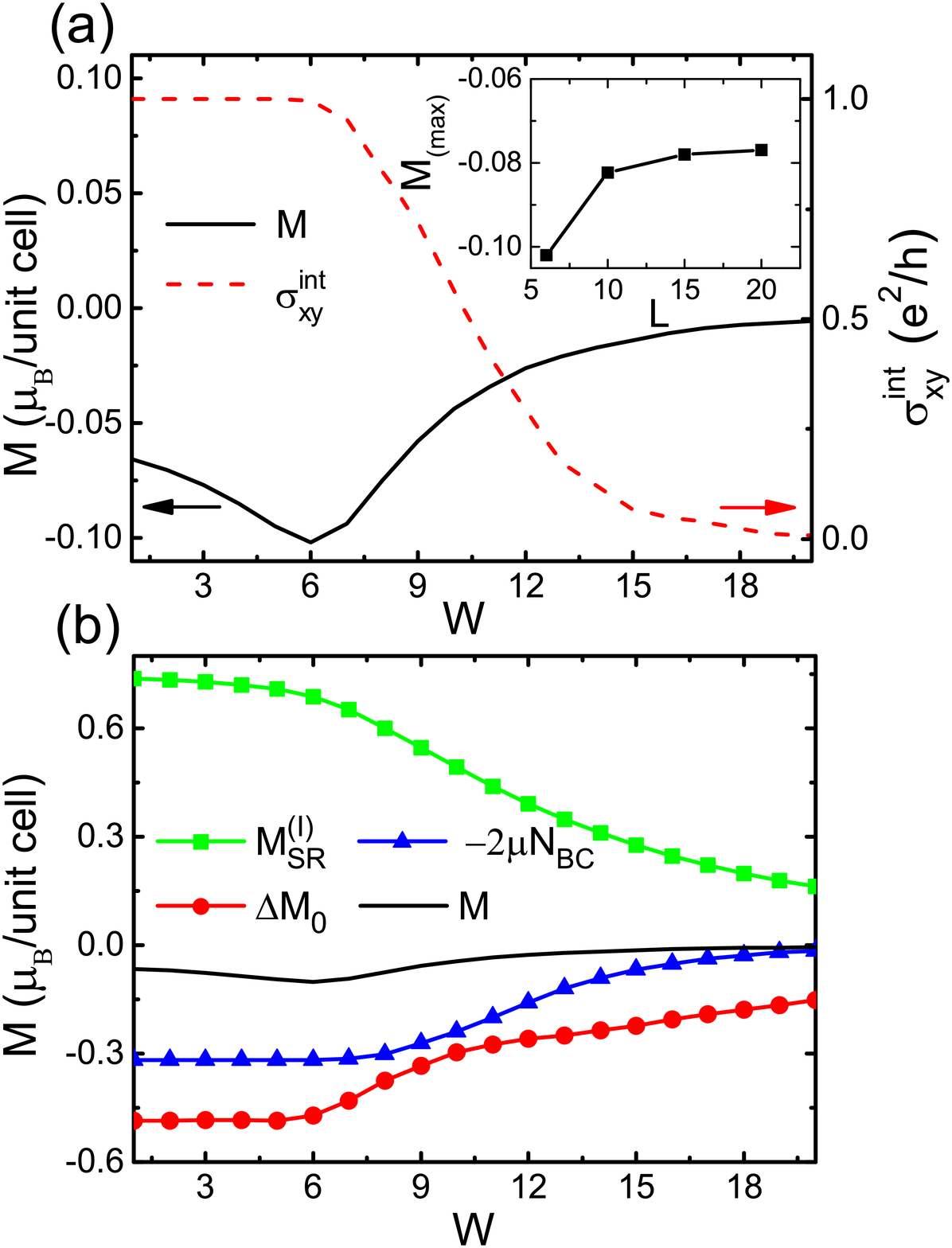}
\caption{ Similar to Fig. \ref{fgh-C=0}, but corresponding to the topologically nontrivial case shown in Fig. \ref{disper} (b). }
\label{fgh-C=1}
\end{figure}

Fig. \ref{fgh-C=1} focuses on the developments of the Chern \emph{insulator}, i.e., by fixing fermi energy $\mu=1$ fixed at the gap center of the clean limit. Fig. \ref{fgh-C=1} (a) is the OM $M$ (black solid line) and AHC $\sigma^{\mathrm{int}}_{xy}$ (red dashed line) under increasing disorder. Different from most of the \emph{band} OM with monotonic dependence on $W$, now there is a peak of $|M|$ at an intermediate disorder $W\sim 6$, just when the AHC plateau starts to collapse. The size dependence of this peak value [inset of Fig. \ref{fgh-C=1} (a)] slows down after $L>10$, so we believe the peak value of $M$ will also approach a stable one in the thermodynamic limit $L\rightarrow \infty$. Similar to the previous normal insulator case, this peak is closely related to the emergence of a disorder induced metal state with non-quantized Hall conductance\cite{YY2013,TopologicalMetal2,TopologicalMetal}. Analogous remarkable disorder enhancement of orbital magnetic moment around the collapse of the AHC plateau was also found in bilayer quantum anomalous Hall systems, where it manifests itself as a peak of orbital magnetoelectric coupling\cite{SS2019}. We conjecture that such disorder enhancement of orbital magnetic motion\cite{YY2013,TopologicalMetal2,TopologicalMetal,SS2019} is an indication of a ``topological metal''\cite{TopologicalMetal2,TopologicalMetal} before localization. This also offers an option to finding materials with remarkable OM.

The developments of corresponding three components is shown in
Fig. \ref{fgh-C=1} (b). The most noticeable difference from the trivial state in Fig. \ref{fgh-C=0} (b), is the remarkable contribution from the Berry curvature term $-2\mu N_{\mathrm{BC}}$ (blue triangles) due to nonzero Chern insulator.

\section{Summary}

In summary, the OM $M$ in two dimension under disorder is studied, based on the two-band Haldane model whose Chern number can be conveniently tuned.

Starting from a normal insulator, disorder will bring two bands together, and induce a ``topological metal'' with nonzero AHC $\sigma_{xy}^{\mathrm{int}}$ in the band touching region. This metallic state corresponds to a disorder induced peak of OM. On the other hand, the OM profiles associated with both bands are shifting along the energy axis, consistent with previous analytical predictions. Besides, our numerical simulations show that there is always a magnitude reduction accompanying with the shifts, reflecting the localization tendency of orbital motions.

Starting from a Chern insulator with a fixed fermi energy in the gap of clean limit, there also appears an $|M|$ peak with increasing disorder, almost simultaneous with the collapse of the quantized Chern number. As for the band OM, it is greatly influenced by the contribution from chiral edge states pinned at the bulk gap, and is therefore deviated from the energy renormalization picture derived from the self-consistent T-matrix approximation.

\ack
We thank Prof. G. B. Zhu (Heze University) for beneficial discussions. This work was supported by National Natural Science Foundation of China under Grants No. 11774336 and No. 61427901. Y.-Y.Z. was also supported by the Starting Research Fund from Guangzhou University.

\end{document}